\documentclass[twocolumn,showpacs,preprintnumbers,amsmath,amssymb]{revtex4}

\usepackage{graphicx}
\usepackage{dcolumn}
\usepackage{bm}

\begin{document}

\preprint{APS/123-QED}

\title{Effects of chemical substitution on transport properties of Bi-based high temperature superconductors}
\author{Abebe Kebede}
\email{abkebede@gmail.com}

\affiliation{Department of Physics}
\address{North Carolina Agricultural and Technical State
University} \address{1601 E. Market St., Greensboro, NC 27411 USA}

\date{\today}
\begin{abstract}
Abstract: This report describes some transport and magnetic
properties of doped Bi2212 and Bi2223 superconducting whiskers.
These materials have advantages over polycrystalline sample as well
as large bulk crystals in that they provide narrow transitions in
resistance versus temperature as well as in magnetization versus
temperature curves. In addition they are easy to grow and and have
short oxygen annealing times.  Here Data on transport and magnetic
properties of these superconducting whiskers are presented.

\end{abstract}


\maketitle
\section{INTRODUCTION}
One characteristic of the copper-oxide superconductors is the close
relation between magnetism and superconductivity. In particular it
has been found that transition metal substitutions for  Cu in
$YBa_2Cu_3O_7$(YBCO) and its rare earth analogues, dramatically
affect the transition temperature $T_c$. On the other hand
substitutions of rare earth impurities metals for Y  were found to
have little or no effect on $T_c$\cite{REF1}.

In the case of Bi-based high $T_c$ superconductors of the form
$Bi_2$$(SrCa)_{2+n}$($Cu_{1-x}$$M_x$)$_{1+n}$$O_y$, where M
represents transition metals such as Zn and Ni  as well as rare
earth metals such as Gd, it is found out that both groups depress
$T_c$ in sharp contrast to what is observed in the YBCO
system\cite{REF1}

$Bi_2$$Sr_2$$Ca_2)$($Cu_{1-x}$$M_x$)$_3$$O_y$ (Bi2223) and
$Bi_2$$Sr_2$$Ca_1)$($Cu_{1-x}$$M_x$)$_2$$O_y$  (Bi2212) whiskers
were prepared at various doping levels\cite{REF2}.  Starting powders
were mixed and melted at 1200C for 30 min.  The molten material was
then quenched to room temperature in about two seconds to form a
glassy material.  The whiskers were grown by annealing the resulting
glassy material in an oxygen environment at 840 C for five days and
followed by cooling to room temperature.

\section{Experimental Results}

Typical dimensions of the whiskers used in our measurements are 2
$\mu m$ to 10$\mu m$(a axis) X 10$\mu  m$to 100$\mu m$ (b axis) X
2$\mu m$ to 10$\mu m$ (c axis). They have mass up to 100 $\mu g$ to
300 $\mu g $. The electric field and magnetic field configurations
relative to the crystal planes is shown in Figure 1.

\begin{figure}[htb]
\center{\includegraphics[scale=0.4]{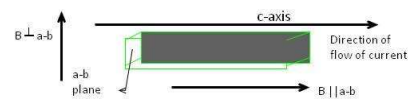}} \caption{Electric
current and magnetic field configuration} \label{Fig}
\end{figure}

The resistance, Meissner effect and critical current measurements
were performed. The  field dependence of critical current density
and resistance versus temperature were also measured. Shown in
Figures 2, is a typical normalized resistance (R/R250K) versus
temperature of Bi2223 with 0.08 Zn concentration measured using the
four probe technique.

\begin{figure}[htb]
\center{\includegraphics[scale=0.4]{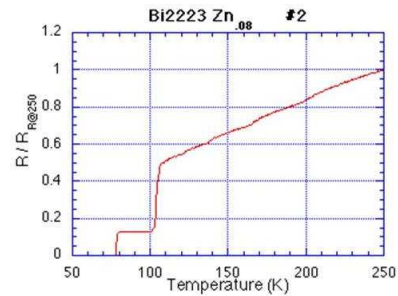}} \caption{Resistance
versus Temperature showing double transition indicating the presence
of both Bi2212 and Bi2223 phases} \label{Fig}
\end{figure}

Our best measurement showed  a transition width of less than 10K for
most samples. These samples undergo double transitions. The first
transition near 100 K and the  second near 80 K; showing the
presence of both Bi2223 and Bi2212 phases in any batch.

We also measured the superconducting transition temperature using a
Quantum design SQUID magnetometer  with the aim of determining  bulk
superconductivity.  Figure 3  shows a typical magnetization curve
for x =0.08 Zn concentration.  For most samples the observed
Meissner effect was only due to the Bi2212 phase with transition
near 80K. The high temperature transitions observed in the
resistance data were due to filamentary conductivity. Figure 4 shows
the transition temperature decreases as the concentration of Zn
increases at the rate $dT_c/dx \sim 250 K/mole Zn$ . This decrease
shows that Zn is a chemical impurity that introduces some
modification of superconductivity as opposed to a magnetic impurity
that would dramatically reduce $T_c$ to Zero.

\begin{figure}[htb]
\center{\includegraphics[scale=0.4]{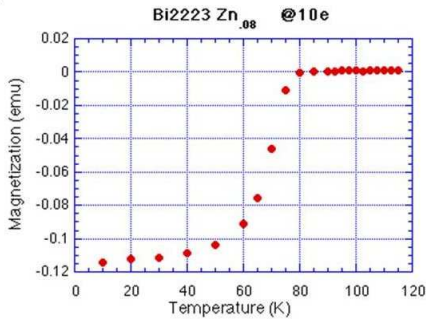}}
\caption{Magnetization versus Temperature at $H=10Oe$ and $x=0.08$}
\label{Fig}
\end{figure}

\begin{figure}[htb]
\center{\includegraphics[scale=0.4]{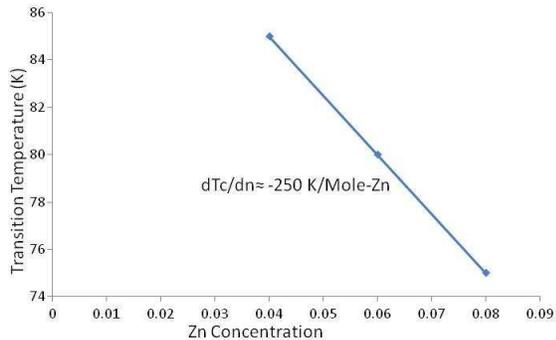}} \caption{$T_c$ versus
Zn concentration showing Zn acts like a chemical impurity rather
than magnetic impurity} \label{Fig}
\end{figure}

We have also measured the resistance of this sample in magnetic
field in the range of 0 $<$ $H$ $<$ 9 $T$  as shown in Figure 5. We
observed that the high temperature transition (110K) became broad.
The transition corresponding to  Bi2212 became too broad to observe.
The magnetic field  ($H_c$) is plotted against the mid point of the
transition to superconductivity in Figure 6.  The rate of decrease
of the critical magnetic field as a function of $T_c$ ( $dH_c/dT_c)
\sim -0.5T/K$

\begin{figure}[htb]
\center{\includegraphics[scale=0.4]{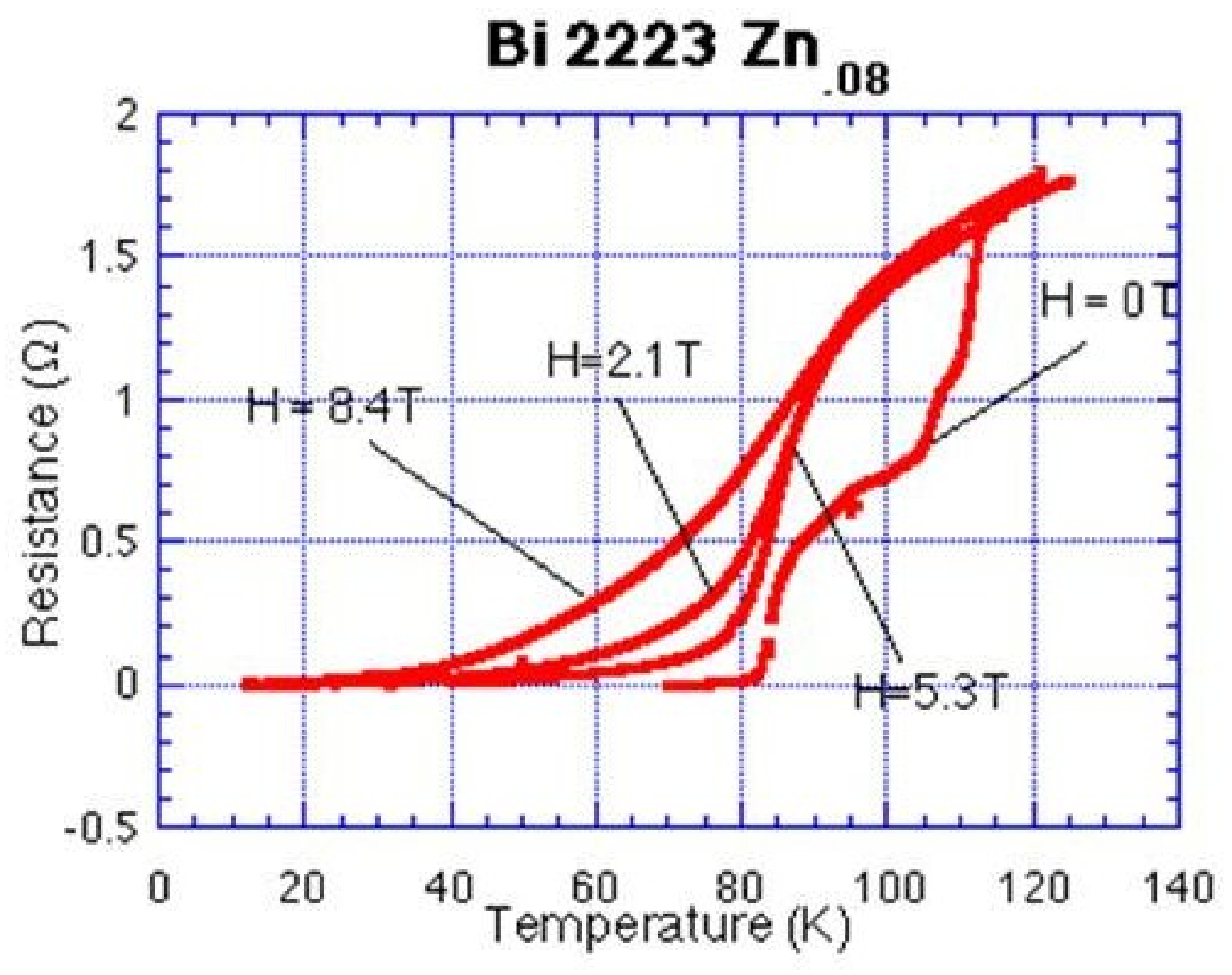}} \caption{Resistance
versus temperature as a function of magnetic field} \label{Fig}
\end{figure}

\begin{figure}[htb]
\center{\includegraphics[scale=0.4]{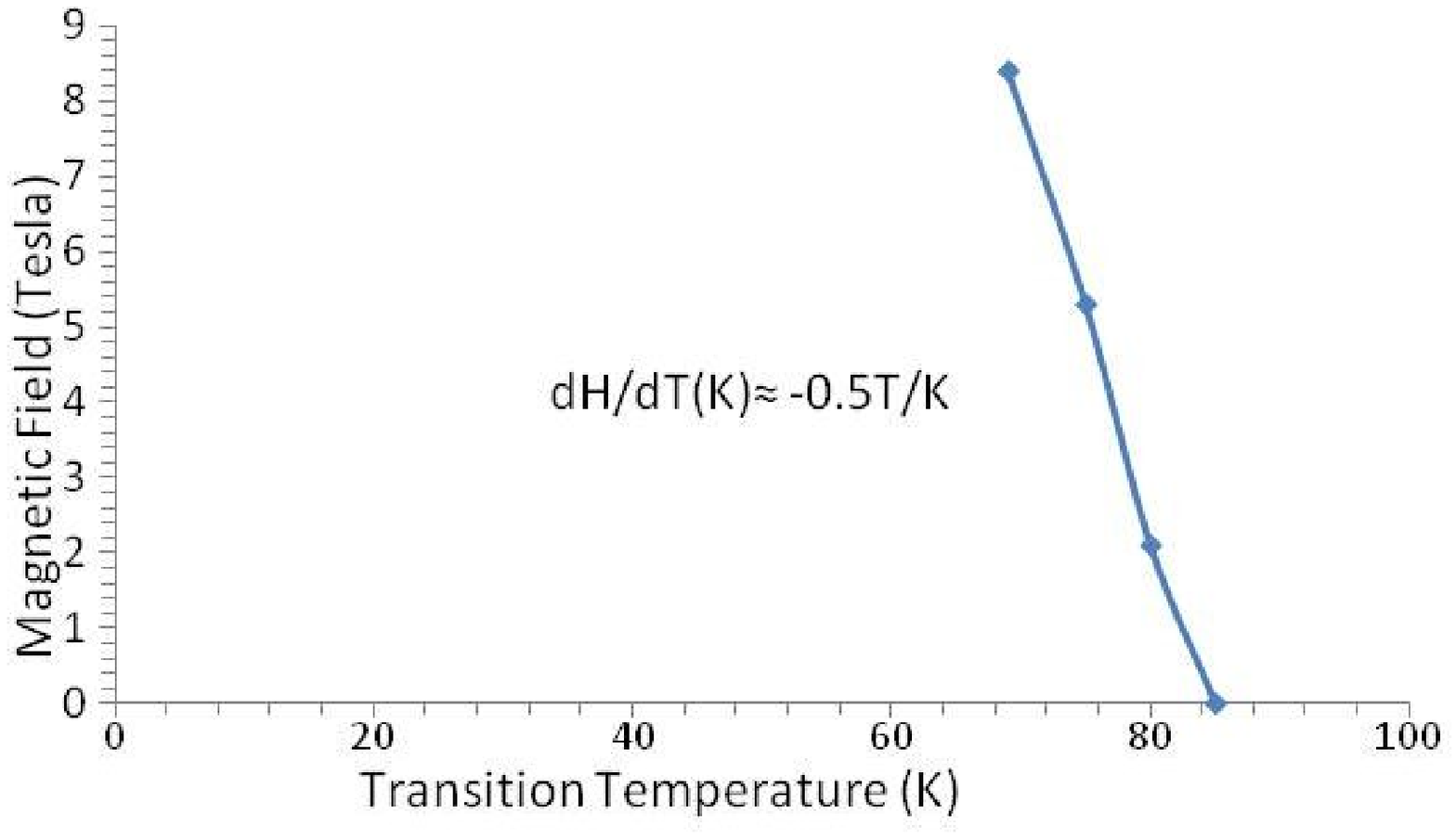}} \caption{Resistance
versus temperature as a function of magnetic field} \label{Fig}
\end{figure}

We measured the superconducting critical currents for the $x=0.06$
Zn in field (0$<$ $H$ $<$ 1 $T$) parallel  and perpendicular to the
c axis. The results are shown in Figures 7 and  8. $J_c$ at 64K in
field parallel to a-b plane, ranges from 1600 $A/cm^2$ (Zero field)
to 600 $A/cm^2$  (1 Tesla). On the other hand the data for the case
of the field parallel to the $c$-axis at 64K, show a rapid decrease
of $J_c$ in field and  it was suppressed to zero using a field of
0.1 T.
\begin{figure}[htb]
\center{\includegraphics[scale=0.4]{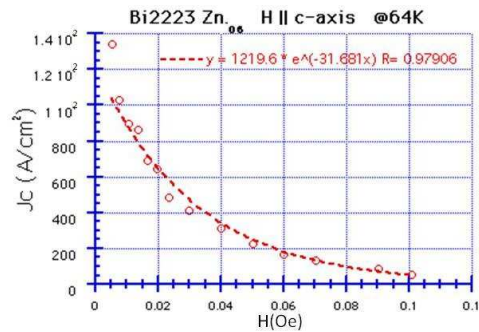}} \caption{$J_c$ is
extracted from Figure 7. The data fit an exponential decay function}
\label{Fig}
\end{figure}

\begin{figure}[htb]
\center{\includegraphics[scale=0.4]{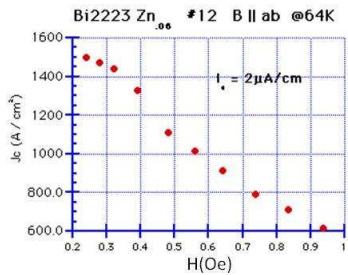}} \caption{$J_c$ when
the filed is parallel to the ab plane. The data fit an a linear
function} \label{Fig}
\end{figure}
We conducted similar $I-V$ measurements when the field is parallel
to the $ab$ plane, the extracted critical density is shown Figure 8.
In this case the current density is much more robust to the applied
magnetic field and it decreases linearly as the field increases.

\section{Conclusions}

In this work we have measured typical transport and magnetic
properties of Zn doped Bi based high temperature superconductors.
Our data indicate that the whiskers are bulk superconductors mostly
composed of the Bi2212 phase. We have shown that the critical
current density is dependent on the crystal direction.  Its value is
higher when the magnetic field is parallel to the $ab$ plane, than
when the later is parallel to the $c$-axis. The critical currents,
obtained at 64K in 0-1T field range, of our Bi2223 and Bi2212
whiskers are small compared to the values for polycrystalline
Ag-sheathed Bi2223 tapes\cite{REF3}. This is an indication that
doping with Zn didn't introduce effective pinning centers. The
dependence of the critical temperature on Zn concentration is
typical of the presence of chemical impurity in the system.

\end{document}